\documentclass[reqno,12pt]{article}
\usepackage{amsmath, amsfonts, amscd, eucal}
\usepackage{latexsym}
\usepackage{amssymb}

\newtheorem{lem}{Lemma}[section]
\newtheorem{prop}{Proposition}[section]

 \numberwithin{equation}{section}

\begin{document}

\title{A Generalized Positive Energy Theorem for Spaces with
Asymptotic SUSY Compactification}\author{Naqing Xie\footnote{{\em Email Address:}
031018008@fudan.edu.cn}\\[0.3cm]
{\normalsize\it Institute of Mathematics, Fudan University,
Shanghai 200433, P.R.China}}
\date{}

 \maketitle \begin{abstract}In this short note, we prove a
generalized positive energy theorem for spaces with asymptotic
SUSY compactification involving non-symmetric data. This work is
motivated by the work of Dai \cite{D1}\cite{D2},
Hertog-Horowitz-Maeda \cite{HHM}, and Zhang \cite{Z}.
\end{abstract}
\pagestyle{plain} {~~~

{\em JGP SC:} Global Differential Geometry, Supersymmetric field
theory

\section{Introduction and Statement of the Result}In 1960, Arnowitt-Deser-Misner  made a
detailed study of isolated gravitational systems from the
Hamiltonian point of view \cite{ADM}. They discovered a conserved
quantity given precisely by an integral and they concluded that
this conserved quantity is the total energy of this isolated
system. Mathematically rigorous proof of the conjecture that the
total energy for asymptotically flat spaces is nonnegative was
firstly given by Schoen and Yau \cite{SY1}\cite{SY2}\cite{SY3}.
Shortly thereafter, Witten raised a simple proof using spinors
from 'spacetime' view \cite{Wi}\cite{PT}. Later, various results
have been established: Bartnik \cite{B} defined the ADM mass for
higher dimensional spin manifolds and generalized this theorem to
that case; Zhang \cite{Z} globally defined the concept of angular
momentum and proved a positive mass theorem involving this
nonsymmetric data which gave an answer to the $120^{th}$ problem
of Yau in his problem section \cite{Y}.

In string theory \cite{CHSW}, our universe is modelled by a ten
dimensional manifold which asymptotically approaches the product
of a flat Minkowski space $M^{3,1}$ with a compact Calabi-Yau
$3$-fold $X$. This is the so-called Calabi-Yau compactification
which motivates the spaces we discuss here. Hertog-Horowitz-Maeda
constructed classical configuration which has regions of negative
energy density as seen from four dimensional perspective
\cite{HHM}. This guides us to revisit the concept of the ADM mass
(or the total energy) in string theory. A positive mass theorem
for such spaces was established by Dai \cite{D1} and its
Lorentzian version was discussed in \cite{D2}.

In this short note, we formulate and prove a generalized positive
energy theorem for spaces with asymptotic SUSY compactification
which involves non-symmetric initial data.

We consider the complete Riemannian manifold $(M^n,
g_{ab},p_{ab})$ with non-symmetric data $p_{ab}$. Suppose $M=M_0
\cup M_\infty$ with $M_0$ compact and $M_\infty \simeq
(\mathbb{R}^k-B_R(0))\times X$ for some $R >0$ and $X$ a compact
simply connected Calabi-Yau manifold. We will call $(M^n,
g_{ab},p_{ab})$ a space with asymptotic SUSY compactification if
the metric on the end $M_\infty$ satisfies the following
asymptotic conditions \begin{equation}g=\overset{\circ}{g}+h, \
\overset{\circ}{g}=g_{\mathbb{R}^k}+g_X,\end{equation}
 \begin{equation}\ h=O(r^{-\tau}),
\ \overset{\circ}{\nabla}h=O(r^{-\tau-1}), \
\overset{\circ}{\nabla}\overset{\circ}{\nabla}h=O(r^{-\tau-2}),\end{equation}
and \begin{equation}p=O(r^{-\tau-1}), \
\overset{\circ}{\nabla}p=O(r^{-\tau-2})\end{equation} where
$p_{ab}$ is an arbitrary two-tensor satisfying $p_{\beta
\alpha}=p_{\beta i}=p_{i \beta}=0$, $\overset{\circ}\nabla$ is the
Levi-Civita connection with respect to $\overset{\circ}g$,
$\tau>0$ is the asymptotic order, $r$ is the Euclidean distance to
a base point, and the index $\alpha, \beta$ run over the compact
factor while the
index $i$ runs over Euclidean part.\\
\indent For such a space $(M^n, g_{ab},p_{ab})$, the total energy
is defined as
\begin{equation}E= \lim_{R \rightarrow \infty} \frac{1}{4 \omega_k
vol(X)} \int_{S_R \times X} (\partial_i g_{ij} - \partial_j g_{aa}
) * dx_j dvol(X),\end{equation}
 and the total momentum is defined
as \begin{equation} P_k = \lim_{R \rightarrow \infty} \frac{1}{4
\omega_k vol(X)} \int_{S_R \times X} 2(p_{kj} - \delta_{kj} p_{ii}
)
* dx_j dvol(X). \end{equation} Here the $*$ operator is the one on
the Euclidean factor, the index $i, j, k$ run over the Euclidean
factor while the index $a$, $b$ run over the
full index of the manifold.\\
\indent We say that $(M^n, g_{ab},p_{ab})$ satisfies the dominant
energy condition if \begin{equation}\label{DCT}\mu \geq
max{\Big\{}\sqrt{\sum_a(\omega_a)^2},
\sqrt{\sum_a(\omega_a+\chi_a)^2}{\Big\}}+\sqrt{\sum_{1 \leq a \leq
n-3}\kappa_a^2}.\end{equation}Here, local energy density is
defined as
\begin{equation}
\mu=\frac{1}{2}(R+(\sum_ap_{aa})^2-\sum_{a,b}p_{ab}^2)
\end{equation}
where $R$ is the scalar curvature, and local momentum densities
are defined as
\begin{equation}
\omega_a=\sum_b(\nabla_bp_{ab}-\nabla_ap_{bb}),
\end{equation}
\begin{equation}
\chi_a=2\sum_b \nabla_b \widetilde{p}_{ba},
\end{equation}
\begin{equation}
\kappa_a^2=\sum_{b,c,d;
c>d>b>a}(\widetilde{p}_{ab}\widetilde{p}_{cd}+\widetilde{p}_{ac}\widetilde{p}_{db}+\widetilde{p}_{ad}\widetilde{p}_{bc})^2,
\end{equation}
where $\widetilde{p}_{ab}=p_{ab}-p_{ba}$.

Our main result is\\[0.5cm]
{\large\bf Main Theorem.} {\em  Let $(M^n, g_{ab},p_{ab})$ be a
complete spin manifold as above and the asymptotic order $\tau>
\frac{k-2}{2}$ and $k \geq 3$. If $(M^n, g_{ab},p_{ab})$ satisfies
the dominant energy condition (\ref{DCT}), then one has
\begin{equation}
E \geq |P|.
\end{equation}
Moreover, if $E=0$ and $k=n$, then the following equation holds on
$M$
$$\sum_{c <
d}(R_{abcd}+p_{ac}p_{bd}-p_{ad}p_{bc})e^ce^d-\sqrt{-1}\sum_c(\nabla_ap_{bc}-\nabla_bp_{ac})e^c$$
\begin{equation}=-\sqrt{-1}(\sum_{c,d; a \neq c \neq d \neq b \neq
a}\nabla_ap_{cd}e^be^ce^d-\sum_{c,d; a \neq c \neq d \neq b \neq
a}\nabla_bp_{cd}e^ae^ce^d)\end{equation}
$$-(\sum_{f,c,d; a \neq f \neq c \neq d \neq b \neq a}p_{cd}p_{af}e^be^fe^ce^d-\sum_{f,c,d; a \neq f \neq c \neq d \neq b \neq a}p_{cd}p_{bf}e^ae^fe^ce^d)$$as an endomorphism of the spinor bundle $S$, where $R_{abcd}$ is the Riemann curvature tensor of the manifold $(M^n,
g_{ab},p_{ab})$.}\\[0.5cm]
{\em\noindent Remarks:}\\
1. This theorem extends without change to the case of $X$ with any
other special holonomy except $Sp(m) \cdot SP(1)$.\\
2. In particular, if the data $p_{ab}$ is symmetric, then this
theorem reduces to the result in \cite{D2}.\\
3. This theorem corresponds to the result in \cite{Z} in the
asymptotically flat case.

\section{The Bochner-Lichnerowicz-Weitzenbock Formula}
Our argument is inspired by Witten \cite{Wi}\cite{PT}. We will
adapt the spinor method \cite{Z}\cite{D1}\cite{D2} to our
situation. The crucial point is that we use the Dirac-Witten
operator $\widetilde{D}$ which is defined in \cite{Z}. Our
positive energy theorem is a consequence of a nice generalized
Bochner-Lichnerowicz-Weitzenbock formula.

Fix a point $p \in M$ and an orthonormal basis $\{e_a\}$ of $T_pM$
such that $(\nabla_ae_b)_p=0$, where $\nabla$ is the Levi-Civita
connection of $M$. Let $\{e^a\}$ be the dual frame. Let $S$ be the
spinor bundle of $M$ with Hermitian metric $<\cdot , \cdot
>$. The connection $\nabla$ of $M$ induces a connection on $S$.
Define the modified connections $\widetilde{\nabla}$ and
$\overline{\nabla}$ on $S$ as
\begin{equation}
\widetilde{\nabla}_a=\nabla_a+\frac{\sqrt{-1}}{2}\sum_{b}p_{ab}e^b,
\end{equation}
\begin{equation}
\overline{\nabla}_a=\nabla_a+\frac{\sqrt{-1}}{2}\sum_{b}p_{ab}e^b-\frac{\sqrt{-1}}{2}\sum_{b,c;a\neq
b\neq c\neq a}p_{bc}e^a e^b e^c.
\end{equation}
Then the Dirac operator $D$ and the Dirac-Witten operator
$\widetilde{D}$ are defined as
\begin{equation}
D=\sum_{a}e^a \nabla_a,
\end{equation}
\begin{equation}
\widetilde{D}=\sum_a e^a \widetilde{\nabla}_a
\end{equation}
respectively. Moreover, we have the following formulae:
\begin{equation}\label{d1}
d(<\phi, \psi>int(e^a)dvol)=(<\widetilde{\nabla}_a\phi, \psi
>+<\phi, (\widetilde{\nabla}_a-\sqrt{-1}\sum_bp_{ab}e^b)\psi>)dvol\end{equation}
\begin{equation}\label{d2}
=(<\overline{\nabla}_a\phi, \psi
>+<\phi, (\overline{\nabla}_a-\sqrt{-1}\sum_bp_{ab}e^b)\psi>)dvol,\end{equation}
\begin{equation}\label{d3}
d(<e^a\phi, \psi>int(e^a)dvol)=(<\widetilde{D}\phi, \psi
>-<\phi, (\widetilde{D}+\sqrt{-1}\sum_ap_{aa})\psi>)dvol.\end{equation}
We denote the adjoint operators by
\begin{equation}\label{ad1}
\widetilde{\nabla}_a^\ast=-\widetilde{\nabla}_a+\sqrt{-1}\sum_bp_{ab}e^b,
\end{equation}
\begin{equation}\label{ad2}
\overline{\nabla}_a^\ast=-\overline{\nabla}_a+\sqrt{-1}\sum_bp_{ab}e^b,
\end{equation}
\begin{equation}\label{ad3}
\widetilde{D}^\ast=\widetilde{D}+\sqrt{-1}\sum_ap_{aa}.\end{equation}
Now we recall two nice formulae in \cite{Z}.
\begin{prop}
One has \begin{equation}\label{l1}
\widetilde{D}^\ast\widetilde{D}=\overline{\nabla}^\ast\overline{\nabla}+\frac{1}{2}(\mu+\sqrt{-1}\sum_b\omega_be^b)+\frac{1}{2}\mathcal{F},
\end{equation}
\begin{equation}\label{l2}
\widetilde{D}\widetilde{D}^\ast=\overline{\nabla}^{}
\overline{\nabla}^\ast+\frac{1}{2}(\mu-\sqrt{-1}\sum_b(\omega_b+\chi_b)e^b)-\frac{1}{2}\mathcal{F}\end{equation}
where $\mathcal{F}=\sum_{a \neq b \neq c \neq d \neq
a}p_{ab}p_{cd}e^a e^b e^c e^d$.\end{prop} We are going to derive
the integral form of the generalized
Bochner-Lichnerowicz-Weitzenbock formula.
\begin{lem}
One has $$\int_{\partial M}<\phi,
\overline{\nabla}_a\phi+e^a\widetilde{D}\phi>int(
e^a)dvol(g)=\int_M|\overline{\nabla}\phi|^2+\frac{1}{2}<\phi,
(\mu+\sqrt{-1}\sum_a
\omega_ae^a)\phi>$$\begin{equation}\label{BLW}+\int_M\frac{1}{2}<\phi,
\mathcal{F}\phi>-|\widetilde{D}\phi|^2.\end{equation}
\end{lem}
{\bf\noindent Proof.} By (\ref{l1}),
$$RHS=\int_M|\overline{\nabla}\phi|^2+<\phi, \widetilde{D}^\ast\widetilde{D}\phi>-|\widetilde{D}\phi|^2-<\phi, \overline{\nabla}^\ast\overline{\nabla}\phi>$$
$$=\int_{\partial M}<\phi, \overline{\nabla}_a\phi>int(e^a)dvol(g)-\int_{\partial M}<e^a\phi,
\widetilde{D}\phi>int(e^a)dvol(g)=LHS.$$
\section{Manifolds with Parallel Spinors} Recall that the spin
manifold $M=M_0 \cup M_\infty$ with $M_0$ compact and $M_\infty
\simeq (\mathbb{R}^k-B_R(0))\times X$ for some $R >0$. Since $k
\geq 3$ and $X$ is simply connected, the end $M_\infty$ is also
simply connected and therefore has a unique spin structure which
comes from the product of the restriction of the spin structure on
$\mathbb{R}^k$ and the spin structure on $X$. One has the
following result in \cite{Wa}.
\begin{prop}
Let $(M,g)$ be a complete, simply connected, irreducible
Riemannian spin manifold and $N$ be the dimension of parallel
spinors. Then $N >0$ if and only if the holonomy group of $M$ is
one of $SU(m)$, $Sp(m)$, $Spin(7)$, $G_2$.
\end{prop}
{\bf\noindent Remark.} In physics language, manifolds with
parallel spinors are said to be supersymmetric (SUSY).\\[0.5cm]
We denote by $\{e^0_a\}$ the orthonormal basis of
$\overset{\circ}g$ which consists of $\{\frac{\partial}{\partial
x^i}\}$ followed by an orthonormal basis $\{f_\alpha\}$ of $g_X$.
Orthonormaling the orthonormal frame $\{e^0_a\}$ with respect to
$\overset{\circ}g$ yields an orthonormal frame $\{e_a\}$ with
respect to $g$. Moreover,
\begin{equation}
e_a=e^0_a-\frac{1}{2}h_{ab}e^0_b+O(r^{-2\tau}).\end{equation} This
provides a gauge transformation $\mathcal{A}$ of the tangent
bundles on the end $M_\infty$:
$$\mathcal{\mathcal{A}}:SO(\overset{\circ}g) \rightarrow SO(g)$$
$$e_a^0 \mapsto e_a.$$Hence it induces a map from the spinor bundles.

Now we pick a unit norm parallel spinor $\psi_0$ of
$(\mathbb{R}^k, g_{\mathbb{R}^k})$ and a unit parallel spinor
$\psi_1$ of $(X, g_X)$. Then $\phi_0=\mathcal{A}(\psi_0 \otimes
\psi_1)$ defines a spinor of $M_\infty$. We extend $\phi_0$
smoothly inside and note that
\begin{equation}\label{L2}
\nabla\phi_0=O(r^{-\tau-1})\end{equation} which is a consequence
of an asymptotic formula in \cite{D1}.
\section{Fibred Boundary Calculus and the Dirac-Witten Equation}
In this section, we will use the fibred boundary calculus of
Melrose-Mazzeo \cite{MM} to solve the Dirac-Witten equation. The
argument is following Dai's \cite{D1}.

Let $\overline{M}$ be a smooth compact manifold with boundary and
suppose that $x$ is a boundary defining function such that $x$
vanishes on $\partial \overline{M}$ and $dx \neq 0$ there. Assume
further that the boundary $\partial \overline{M}$ comes with a
fibration structure $F \rightarrow \partial \overline{M}
\overset{\pi}{\rightarrow} B$ with fiber $F$. Then the metric $g$
is called a fibred boundary metric if in a neighborhood of the
boundary $\partial \overline{M}$, the metric $g$ takes the form
\begin{equation}
g=\frac{dx^2}{x^4}+\frac{\pi^\ast(g_B)}{x^2}+g_F
\end{equation}
where $g_B$ is a metric on the base $B$ and $g_F$ is a family of
fiberwise metrics.

In the setting of spaces with asymptotic SUSY compactification,
the change of variable $x=\frac{1}{r}$ gives a trivial fibration
$S^{k-1}\times X$.

Sometimes we use the notation $M$ and $\overline{M}$
interchangeably. For a manifold with boundary. we introduce two
Lie algebras of vector fields:\\
$\bullet$ \ b-vector fields \begin{equation}\mathcal{V}_b(M):=\{V
\ | \ V \mbox{\ tangent to the boudnary} \ \partial M
\}\end{equation}
and\\
$\bullet$ \ fibred boundary vector fields
\begin{equation}
\mathcal{V}_{fb}(M):=\{V \in \mathcal{V}_b(M)\ | \ V \mbox{\
tangent to the fiber $F$ at } \
\partial M,\ Vx=O(x^2) \} .\end{equation}
The Sobolev space $L^{p,2}(M,S)$ is defined as
\begin{equation}
L^{p,2}(M,S):=\{\phi \in L^2(M,S)\ | \ \nabla_{V_1}\cdots
\nabla_{V_j}\phi \in L^2(M,S), \ \forall j \leq p, \ V_i \in
\mathcal{V}_b(M) \}.\end{equation}

Let $\gamma \in \mathbb{R}$ and we define the space of conormal
sections of order $\gamma$ by
\begin{equation}
\mathcal{A}^\gamma(M,S):=\{\phi \in C^\infty(M,S)\ | \
|\nabla_{V_1}\cdots \nabla_{V_j}\phi|\leq Cx^\gamma, \ \forall j,
\ V_i \in \mathcal{V}_b(M) \},\end{equation} and the subspace of
polyhomogeneous sections by
\begin{equation}
\mathcal{A}^\ast_{phg}(M,S):=\{\phi \in \mathcal{A}^\ast(M,S) \ |
\ \phi \sim \sum_{Re\gamma_j
\rightarrow\infty}\sum_{k=0}^{N_j}\psi_{jk}x^{\gamma_j}(\log x)^k,
\ \psi_{jk} \in C^{\infty}(\partial M, S) \}.\end{equation} These
expansions are meant in the usual asymptotic sense as $x
\rightarrow 0$ and hold along with all derivatives. The
superscript $\ast$ may be replaced by an index set $I$ containing
all pairs $(\gamma_j,N_j)$ which appear in this expansion.

Denote by $\Pi_0: L^2(M,S) \rightarrow KerD_F$ the natural
orthogonal projector and let $\Pi_{\perp}:=Id-\Pi_0$.

The following proposition is a summary of the results in
\cite{HHMa} (See also \cite{D1}, Theorem 3.1).
\begin{prop} Suppose that $a$ is not an indicial root of
$\Pi_0x^{-1}D\Pi_0$. Then
$$D: \ x^aL^{1,2}(M,S) \rightarrow
x^{a+1}\Pi_0L^2(M,S)\oplus x^a\Pi_\perp L^2(M,S)$$is Fredholm. If
$D\phi=0$ for $\phi\in x^aL^2(M,S)$, then $\phi$ is
polyhomogeneous with exponents in its expansion determined by the
indicial roots of $\Pi_0x^{-1}D\Pi_0$ and truncated at $a$. If
$D\xi=\psi$ for $\psi \in \mathcal{A}^a(M,S)$ and $\xi\in
x^{c-1}\Pi_0L^{1,2}(M,S)\oplus x^c\Pi_\perp L^{1,2}(M,S)$ and
$c<a$, then $\xi \in
\Pi_0\mathcal{A}^I_{phg}(M,S)+\mathcal{A}^a(M,S)$.
\end{prop}
{\noindent\bf Remarks.}\\
1. Strictly speaking, only the metric $\overset{\circ}{g}$ is a
fibred boundary metric. However, it is easy to see that the
results generalize to the metric $g$ (see \cite{D1}). The metric
perturbation produces only a lower order term.\\
2. In our situation, note that
$\widetilde{D}=D+\frac{\sqrt{-1}}{2}\sum_{a,b}p_{ab}e^ae^b=D+O(r^{-\tau-1})$.
It follows from the decay condition of the initial data $p_{ab}$
that the Dirac-Witten operator $\widetilde{D}$ is also a Fredholm
operator from $x^aL^{1,2}(M,S)$
to $x^{a+1}\Pi_0L^2(M,S)\oplus x^a\Pi_\perp L^2(M,S)$.\\
3. The precise forms of these results for the Dirac-Witten
operators $\widetilde{D}$ and $\widetilde{D}^\ast$ are somewhat
different, but one still has the regularity property.\\
4. For the precise definition of the indicial root, we refer to
\cite{MM}\cite{HHMa}. For our purpose, we only note that the set
of indicial roots is discrete.

To prove that the Dirac-Witten operator $\widetilde{D}$ is an
isomorphism under certain conditions, we need the following lemma
inspired by \cite{PT} and \cite{Z}.
\begin{lem}\label{extend}Suppose $(M^n,
g_{ab},p_{ab})$ is a complete spin manifold as above and the
spinor $\phi$ satisfying either $\overline{\nabla}\phi=0$ or
$\overline{\nabla}^\ast\phi=0$. If  $\ \lim_{r \rightarrow
\infty}\phi=0$, then $\phi=0$.
\end{lem}
{\noindent\bf Proof.} By the assumptions, we have
$|d|\phi|^2|=2|<Re\nabla\phi,\phi>|\leq C |p||\phi|^2$ where $C$
is some constant. This implies $|d\log|\phi||\leq C r^{-\tau-1} $
outside a compact set. Fix a point $(r_0, y_0)$ and integrate
along a path from $(r_0,y_0)$ with respect to $r$. Then one has
$$|\phi(r,y_0)| \geq |\phi (r_0,
y_0)|e^{C(r_0^{-\tau}-r^{-\tau})}.$$ Taking $r \rightarrow \infty
$ or taking $(r,y_0)$ to be the zero of $\phi$, we get $\phi(r_0,
y_0)=0$. Hence $\phi=0$ when $r$ is large enough. It follows from
the unique continuation property that $\phi=0$ since $\phi$
satisfies the Dirac-Witten equation. We complete the proof of this
lemma.

\begin{lem}\label{iso}
If the dominant energy condition (\ref{DCT}) holds and
$a>\frac{k-2}{2}$ is not an indicial root, then
$$\widetilde{D} : \ x^aL^{1,2}(M,S) \rightarrow
x^{a+1}\Pi_0L^2(M,S)\oplus x^a\Pi_\perp L^2(M,S)$$ is an
isomorphism.
\end{lem}
{\noindent\bf Proof.} The argument here is similar to Dai's (see
\cite{D1}, Section 3). We first see that $\widetilde{D}$ is
injective. If $\phi \in Ker \widetilde{D} \subset
x^aL^{1,2}(M,S)$, then by elliptic regularity, $\phi \in
\mathcal{A}^a_{phg}(M,S)$. By the Weitzenbock formula (\ref{BLW}),
$$\int_\Omega\{|\overline{\nabla}\phi|^2+\frac{1}{2}<\phi,
(\mu+\sqrt{-1}\sum_a \omega_ae^a)\phi>+\frac{1}{2}<\phi,
\mathcal{F}\phi>\}dvol$$
$$=\int_{\partial \Omega}<\phi,
\overline{\nabla}_a\phi+e^a\widetilde{D}\phi>int(e^a)dvol.$$ By
taking $\Omega$ so that $\partial \Omega=S_r\times X$ and $r
\rightarrow \infty$ we see that the right hand side of the above
equality tends to zero since $\phi \in \mathcal{A}^a_{phg}(M,S)$
and $a>\frac{k-2}{2}$. It follows from the dominant energy
condition (\ref{DCT}) that $\overline{\nabla}\phi=0$ and hence
$\phi=0$ by Lemma \ref{extend}.

The same argument as above applies to the adjoint operator
$\widetilde{D}^\ast$. By the Fredholm property, the surjectivity
of $\widetilde{D}$ follows from the injectivity of
$\widetilde{D}^\ast$ which is a consequence of the Weitzenbock
formula (\ref{l2}) as well as Lemma \ref{extend}. This proves the
lemma.

Now we are ready to solve the Dirac-Witten equation.
\begin{lem}\label{Dirac}
There exists a smooth spinor $\phi$ such that
$\widetilde{D}\phi=0$ and $\phi=\phi_0+O(r^{-\tau})$.\end{lem}
{\noindent\bf Proof.} We construct the wanted spinor by setting
$\phi=\phi_0+\xi$ and solve
$\widetilde{D}\xi=-\widetilde{D}\phi_0=O(r^{-\tau-1})$. By Lemma
\ref{iso}, adjusting $\tau$ slightly if necessary so that it is
not one of the indicial root, we have a solution
$\xi=O(r^{-\tau})$.

\section{Proof of the Main Theorem}
\begin{lem}
If a spinor $\phi$ is asymptotic to $\phi_0$:
$\phi=\phi_0+O(r^{-\tau})$, then one has
\begin{equation}\label{final} \lim_{R \rightarrow \infty }\int_{S_R \times
X}<\phi, \overline{\nabla}_a\phi+e^a\widetilde{D}\phi>int(
e_a)dvol(g)=\omega_k vol(X) <\phi_0,
E\phi_0+\sqrt{-1}P_idx^i\phi_0>. \end{equation}\end{lem}
{\noindent\bf Proof.}\\
$$\int_{S_R \times X}<\phi,
\overline{\nabla}_a\phi+e^a\widetilde{D}\phi>int( e_a)dvol(g)$$
$$=\int_{S_R \times X}<\phi, \nabla_a+\frac{\sqrt{-1}}{2}\sum_{b}p_{ab}e^b-\frac{\sqrt{-1}}{2}\sum_{b,c;a\neq
b\neq c\neq a}p_{bc}e^a e^b e^c\phi>int(e_a)dvol(g)$$
$$+\int_{S_R \times X}<\phi, e^a\sum_be^b(\nabla_b+\frac{\sqrt{-1}}{2}\sum_{c}p_{bc}e^c)\phi>int(e_a)dvol(g),$$
$$=\int_{S_R \times X}<\phi, \nabla_a\phi+e^aD\phi>int(e_a)dvol(X)$$
\begin{equation}\label{en}+\int_{S_R \times X}<\phi,
\frac{\sqrt{-1}}{2}(\sum_bp_{ab}e^b-\sum_{b,c;a\neq b\neq c\neq
a}p_{bc}e^a e^b e^c+\sum_{b,c}p_{bc}e^a e^b
e^c)\phi>int(e_a)dvol(g).\end{equation} The first term in
(\ref{en}) is computed in \cite{D1} which tends to
$\omega_kvol(X)<\phi_0, E\phi_0>$ as $r \rightarrow \infty$. The
second term is
$$\int_{S_R \times X}<\phi,
\frac{\sqrt{-1}}{2}(\sum_bp_{ab}e^b+(\sum_{a=b;b \neq
c}+\sum_{a=c; b \neq c
}+\sum_{b=c})p_{bc}e^ae^be^c)\phi>int(e_a)dvol(g)
$$
$$
=\int_{S_R \times X}<\phi,
\frac{\sqrt{-1}}{2}(\sum_bp_{ab}e^b+\sum_{b\neq
a}p_{ab}e^ae^ae^b+\sum_{b\neq
a}p_{ba}e^ae^be^a+\sum_bp_{bb}e^ae^be^b)\phi>int(e_a)dvol(g)$$
$$
=\int_{S_R \times X}<\phi,
\frac{\sqrt{-1}}{2}(\sum_bp_{ab}e^b-\sum_{b \neq
a}p_{ab}e^b+\sum_{b\neq
a}p_{ba}e^b-\sum_cp_{cc}e^a)\phi>int(e_a)dvol(g)$$
$$
=\int_{S_R \times X}<\phi,
\frac{\sqrt{-1}}{2}(\sum_{b}p_{ba}e^b-\delta_{ba}p_{cc}e^b)\phi>int(e_a)dvol(g)$$
which goes to $\omega_kvol(X)<\phi_0, \sqrt{-1}P_idx^i\phi_0>$ as
$r \rightarrow \infty$.\\[0.5cm]

{\em\noindent Proof of the main theorem:} Now we are ready to
prove our main result. Note that $\sqrt{-1}P_idx^i$ has
eigenvalues $\pm |P|$. We take $\phi_0$ as the unit eigenspinor of
eigenvalue $-|P|$. It follows from the Weitzenbock formula
(\ref{BLW}) that
$$E \geq |P|.$$ The proof of the second part is the same as in \cite{Z}.

\section*{Acknowledgements}
The author is indebted to Professor Xiao Zhang for sharing his
ideas and for many useful discussions on the topic of positive
mass theorems. He also particularly wishes to thank Professor
Chaohao Gu and Professor Hesheng Hu for their encouragement.
Thanks are also due to Professor Xianzhe Dai. This research work
is supported by Doctoral Foundation of Ministry of Education of
China, No. 20030246001 and Natural Science Foundation of China
(Differential Geometry).

\end{document}